\documentclass[3p,times]{elsarticle}

\usepackage{ecrc}

\volume{00}

\firstpage{1}

\journalname{Nuclear Physics A}

\runauth{}

\jid{nupha}

\usepackage{amssymb}

\usepackage{lineno}

\usepackage[figuresright]{rotating}

\newcommand{\CommentBlock}[1]{}

\newcommand{\PbPb}{Pb--Pb}

\newcommand{\pp}{pp}

\newcommand{\sqrtsNN}{\ensuremath{\sqrt{s_{\mathrm {NN}}}}}
\newcommand{\sqrts}{\ensuremath{\sqrt{s}}}

\newcommand{\rr}{\ensuremath{R}}

\newcommand{\gev}{GeV/$c$}

\newcommand{\pT}{\ensuremath{p_\mathrm{T}}}

\newcommand{\kzerol}{\ensuremath{\mathrm{K}^{0}_\mathrm{L}}}

\newcommand{\dxidpt}{\ensuremath$\xi=\ln(p_{\rm{T}}^{\rm{jet}}/p_{\rm{T}}^{\rm{track}})$}
\newcommand{\dNdpT}{dN/d\pT}
\newcommand{\dNdxi}{\ensuremath{\rm{d}N/\mathrm{d} \xi}}

\begin{document}

\begin{frontmatter}



\dochead{}

\title{Measurements of the inclusive jet cross section and jet fragmentation in \pp\ collisions with the ALICE experiment at the LHC}


\author{Rongrong Ma (for the ALICE Collaboration)}
\address{Yale University, New Haven, CT 06520, USA}
\address{E-mail: rongrong.ma@yale.edu}

\begin{abstract}
Jet reconstruction and jet fragmentation variables provide important information to study the interaction between hard scattered partons and the Quark-Gluon Plasma. This paper presents the measurement of the inclusive cross section for fully reconstructed jets in pp collisions at \sqrts\ = 2.76 TeV, which provides an essential reference for jet measurements in Pb-Pb collisions at the same \sqrtsNN\ . In addition, we report jet fragmentation measurements for charged particle jets in pp collisions at \sqrts\ = 7 TeV. These measurements utilize the ALICE central barrel tracking system to detect charged particles with good efficiency above 150 MeV/c, together with the Electromagnetic Calorimeter (EMCal). The jet cross section and fragmentation measurements are compared to theoretical calculations and Monte Carlo generators.
\end{abstract}


\end{frontmatter}


\section{Introduction}
\label{sec:intro}
Jet measurements play an important role in understanding the properties of the strongly-coupled medium created in ultra-relativistic heavy-ion collisions. To quantify the "jet-quenching" phenomena \cite{Majumder:2010qh} in the medium, reference measurements from \pp\ collisions are required. We report measurements of the inclusive differential jet cross section in \pp\ collisions at \sqrts\ = 2.76 TeV and jet fragmentation in \pp\ collisions at \sqrts\ = 7 TeV with the ALICE experiment. Inclusive jet cross sections and their ratio at different values of \rr\ can be calculated using perturbative Quantum Chromodynamics (pQCD) at Next-to-Leading-Order (NLO) \cite{SoyezRatio,Frixione:1997np}, and compared to data. Good agreement is observed. For the jet fragmentation measurement, PYTHIA6 \cite{PYTHIA} (Perugia-0) reproduces data very well in the kinematic range considered.

\section{Data set and Detector}
\label{sec:data}
The data presented here were recorded by the ALICE experiment \cite{ALICE} for pp collisions at  \sqrts\ = 7 TeV in 2010 and \sqrts\ = 2.76 TeV in 2011. Several trigger detectors were used: the VZERO, consisting of segmented scintillator detectors covering full azimuth over $2.8 < \eta < 5.1$ (VZEROA) and $-3.7 < \eta < -1.7$ (VZEROC); the SPD, a highly granular silicon pixel detector close to the beam pipe, and the Electromagnetic Calorimeter (EMCal) covering 100 degrees in azimuth and $|\eta|<0.7$. Both analyses utilized Minimum Bias (MB) triggered events, which required at least one hit in any of VZEROA, VZEROC and SPD, in coincidence with the presence of a bunch crossing. In addition, the cross section analysis used the EMCal triggered events, which required that the MB trigger condition was fulfilled and at least one EMCal ``single shower" sum, a fast sum of energy in groups of $4\times4$ ($\eta\times\varphi$) adjacent EMCal towers each covering an angular region $\Delta\eta\times\Delta\varphi=0.014\times0.014$, exceeded a nominal threshold energy of 3.0 GeV. 

For offline analysis, the input to the FastJet anti-$k_{\rm{T}}$ algorithm \cite{Cacciari:2011ma,Cacciari:2008gp} used for jet reconstruction consisted of charged particle tracks and EMCal clusters. Charged tracks were measured in the ALICE tracking system covering full azimuth within $|\eta|<0.9$. Optimum track momentum resolution and uniform track efficiency were achieved using a hybrid approach: a small fraction of tracks lacking information in the inner tracking system due to local inefficiencies were additionally constrained using the primary vertex. Tracks with measured \pT\ $>$ 0.15 \gev\ were accepted.

EMCal clusters were formed by a clustering algorithm that combined signals from up to $3\times3$ neighboring EMCal towers. Clusters with large apparent energy but anomalously small number of contributing towers, possibly due to the interaction of slow neutrons or highly ionizing particles in the avalanche photodiode of the most energetic tower, were rejected.

\section{Analysis method}
\label{sec:JetRec}
For the jet fragmentation analysis, only charged tracks were used in the jet reconstruction with a resolution parameter of \rr\ = 0.4. We measured the differential momentum distributions, \dNdpT\ and \dNdxi\ (\dxidpt ), of charged particles per jet in bins of charged jet \pT. Jets containing charged tracks above 100 \gev\ were excluded. The leading charged jet  with axis within $|\eta|<0.5$ in each event was used. Reconstructed charged tracks within a distance of $\Delta R=\sqrt{\Delta\eta^{2}+\Delta\varphi^{2}} < 0.4$ with respect to the leading jet axis were used to build the \dNdpT\ and \dNdxi\ distributions. Contributions from the underlying event (UE) not related to the jet fragmentation were estimated via a "perpendicular cone" method in data. In each event, a cone of \rr\ = 0.4 was placed perpendicular to the leading jet axis in the azimuthal direction, and all the tracks within this cone were used to estimate the UE contribution to the signal. Another source of contamination were secondaries, predominantly produced by weak decay of strange particles, photon conversions and the products of hadronic interactions in the detector material. These contributions were estimated from simulations using PYTHIA6 (Perugia-0) event generator followed by detailed GEANT3 \cite{GEANT3} transport and detector response simulation, and were subtracted from data. A data-driven method was utilized to account for the low strangeness yield in the simulation.

The detector effects of tracking efficiency and track momentum resolution can not only shift and smear the measured track distributions inside of jets, but also affect the charged jet reconstruction.  In general, the tracking momentum resolution effectively smeared the jet energy, and the tracking efficiency caused a bin migration in jet \pT. These detector effects were corrected via a bin-by-bin technique based on the same simulation sample mentioned above. The correction factors were extracted by comparing the \dNdpT\ and \dNdxi\ distributions in the same nominal charged jet \pT\ bin on detector level and particle level. In the \pT\ regions considered in this analysis, the correction factor was typically 10-20\%.

In the jet cross section analysis, both charged tracks and EMCal clusters were fed into the jet finding algorithm with two resolution parameters of \rr\ = 0.2 and \rr\ = 0.4. The same charged track selection criteria were used as in the fragmentation analysis. Since the charged hadrons also deposited energy in the EMCal, their contribution to cluster energy should be subtracted to avoid double-counting of energy in the jet measurement. Charged tracks were propagated to the EMCal and matched to the closest cluster within $\Delta\eta = 0.015$ and $\Delta\varphi = 0.03$. The energy of the matched cluster was corrected by removing 100\% of the sum of all associated charged track momenta up to the cluster energy itself. The algorithm potentially over-subtracted cluster energy if the cluster had additional contributions from either neutral particles or undetected charged particles. This residual effect was estimated from PYTHIA+GEANT simulations to be less than 5\% and corrected in the final bin-by-bin correction. Jets fully contained in the EMCal acceptance were accepted.

The jet bias imposed by the EMCal trigger needed to be corrected in order to measure the jet cross section with these data. The trigger efficiency of jets was estimated using a data-driven approach incorporating simulation. Jet events were simulated at the detector level using PYTHIA6 (Perugia-2011)  and GEANT3. Each simulated EMCal cluster was accepted by the trigger with a probability equal to the measured cluster trigger efficiency in data, taking into account the local variations in trigger efficiency. A simulated event was accepted if at least one EMCal cluster in this event was accepted by the trigger. The jet trigger efficiency was determined by comparing the inclusive jet spectrum for the triggered and MB populations in the simulation.

Jets were corrected to the particle level, without accounting for hadronization effects, utilizing a bin-by-bin technique based on simulation. The correction factors were determined by comparing the inclusive jet spectra on detector level and particle level. The main contributions to the correction factors were effects that either shifted or smeared the jet energy. The major sources to the jet energy shift were unmeasured neutron and \kzerol\ , tracking efficiency as well as the residual corrections of charged energy double counting. The jet energy smearing mainly originated from the intrinsic detector resolution and the event-by-event fluctuation of the jet energy shift. The underlying event contribution was not subtracted, but its effect decreased as jet \pT\ increased, and was estimated to be $\sim20\%$ for \rr\ = 0.4 jets and $\sim 5\%$ for \rr\ = 0.2 jets in the lowest jet \pT\ bin. 

\section{Results}
\label{sec:results}

\begin{figure}[htbp]
\centering
\includegraphics[width=0.49\textwidth]{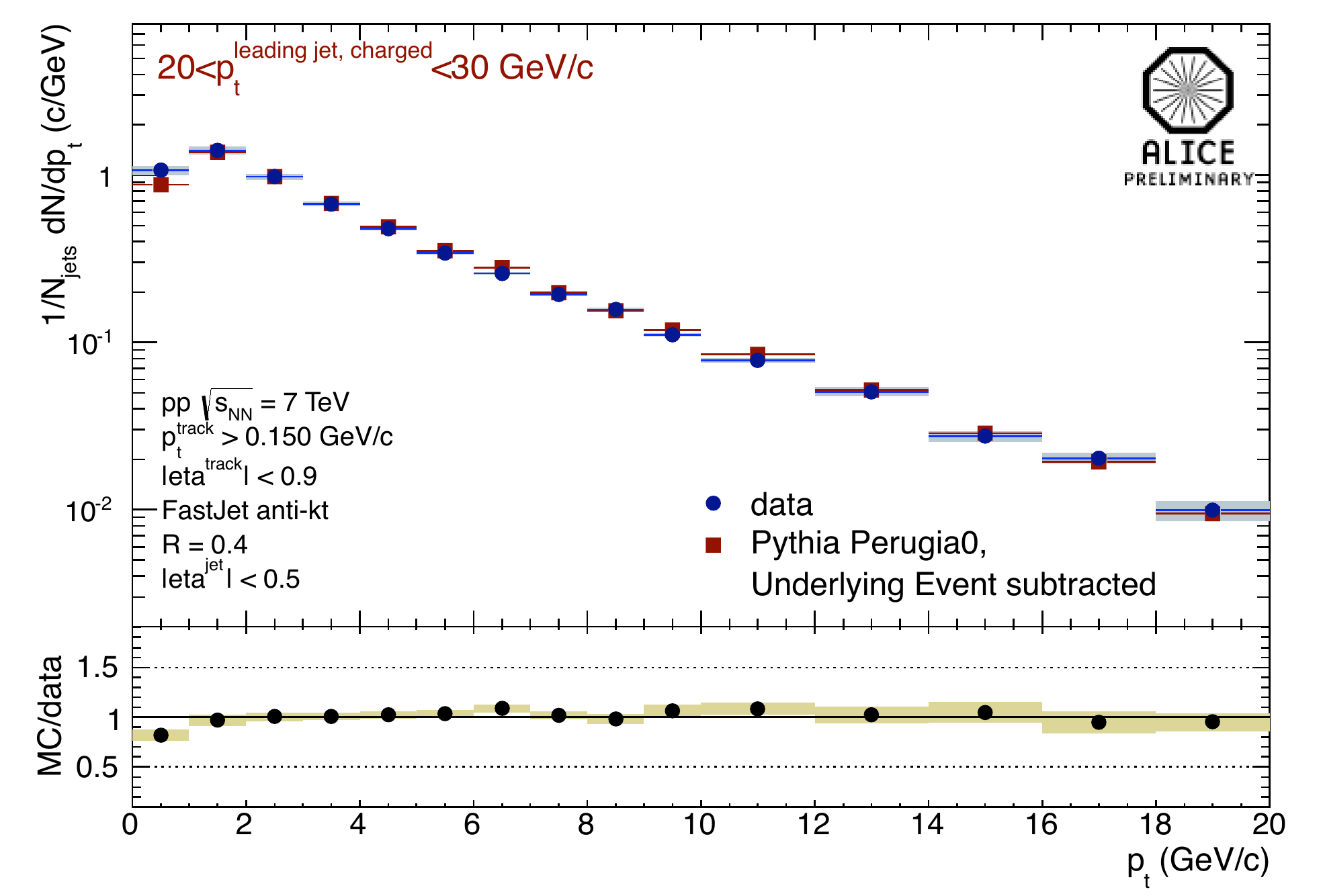}
\includegraphics[width=0.49\textwidth]{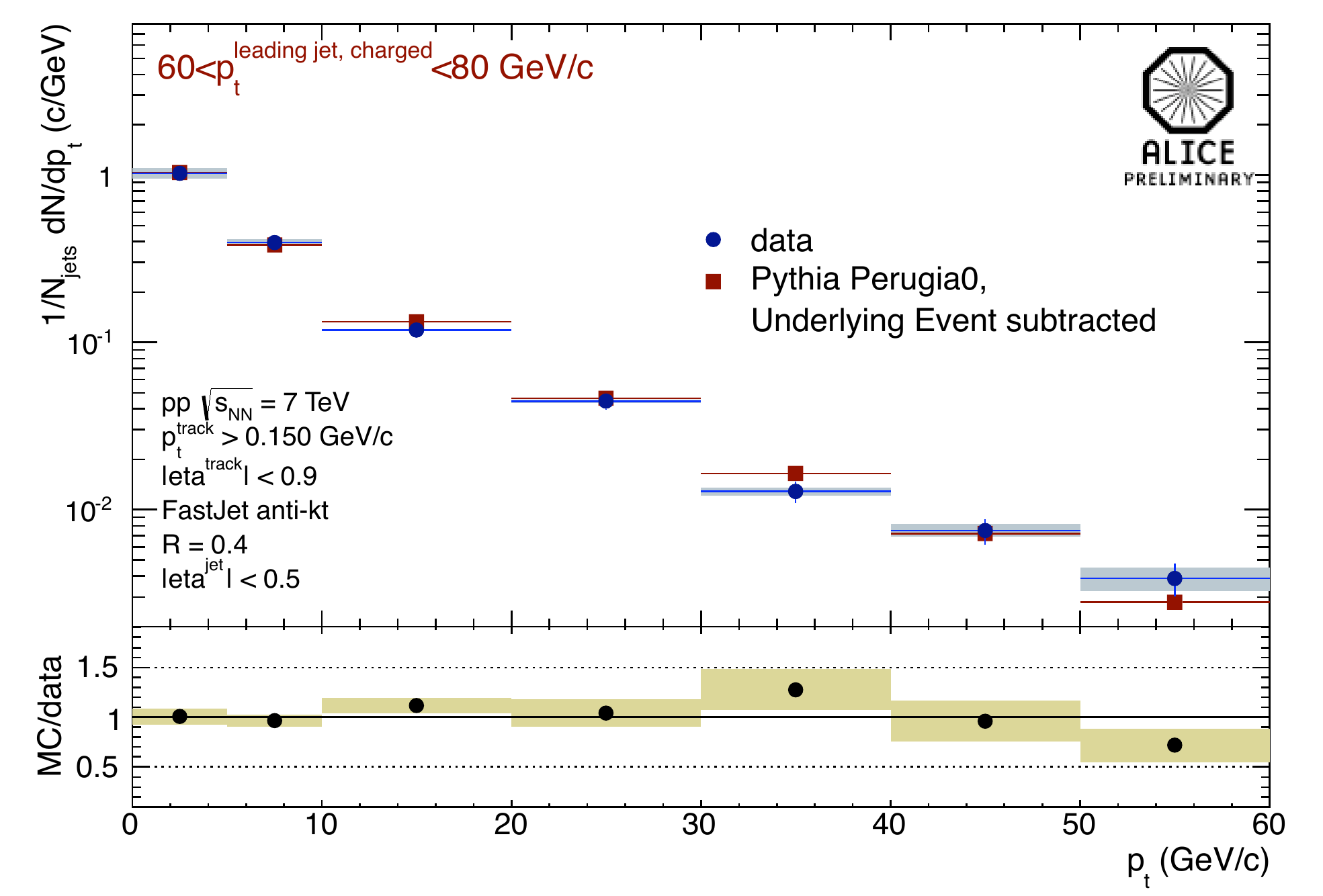}
\caption{Upper panels: \dNdpT\ distributions of charged particles within charged jets in two different charged jet \pT\ bins: 20-30 \gev\ (left) and 60-80 \gev\ (right). Data points (blue circles) are overlaid with PYTHIA6 (Perugia-0) predictions (red squares). Lower panels: ratio of PYTHIA6 predictions to data.  
\label{fig:dNdpTCH}
}
\end{figure}

\begin{figure}[htbp]
\centering
\includegraphics[width=0.49\textwidth]{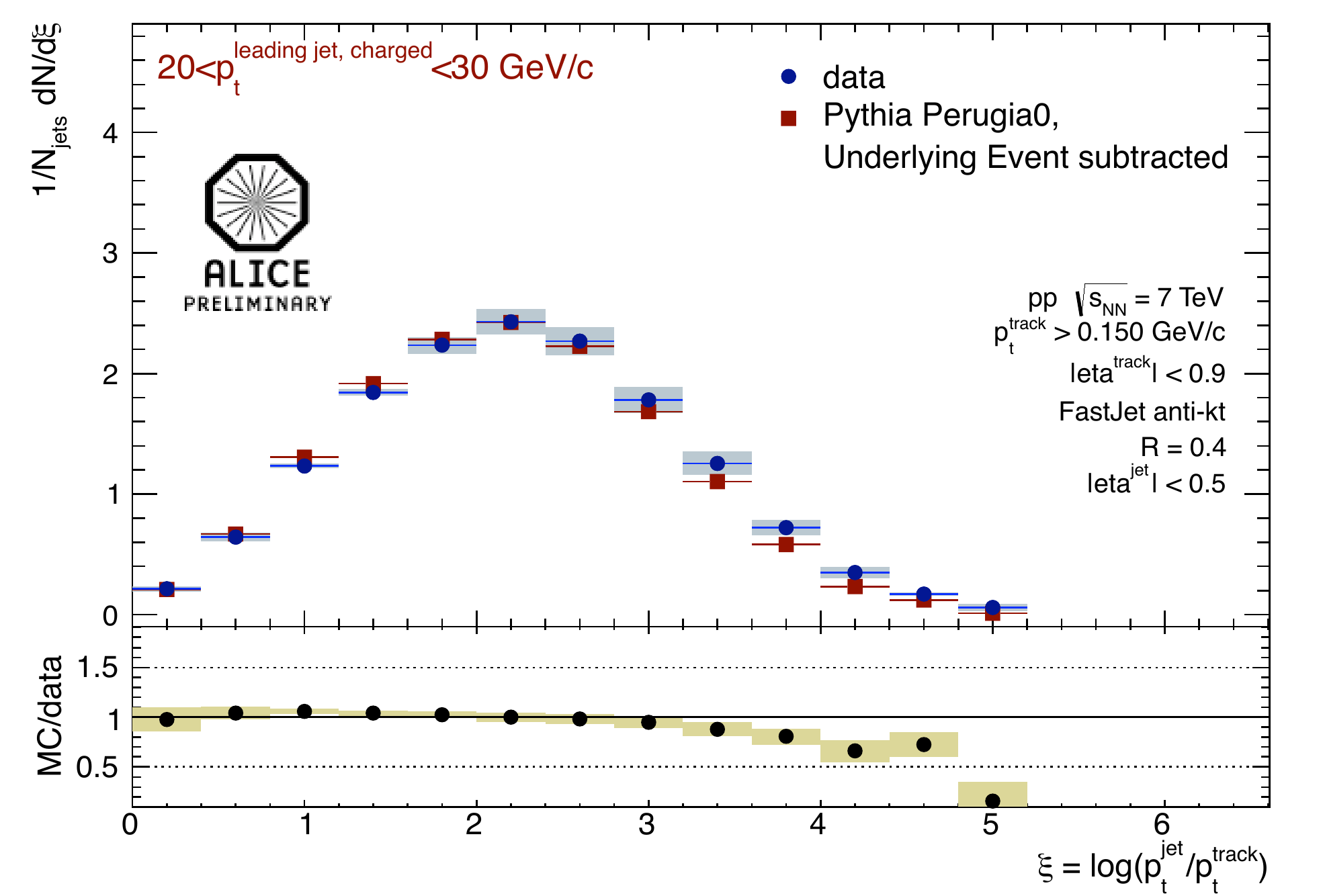}
\includegraphics[width=0.49\textwidth]{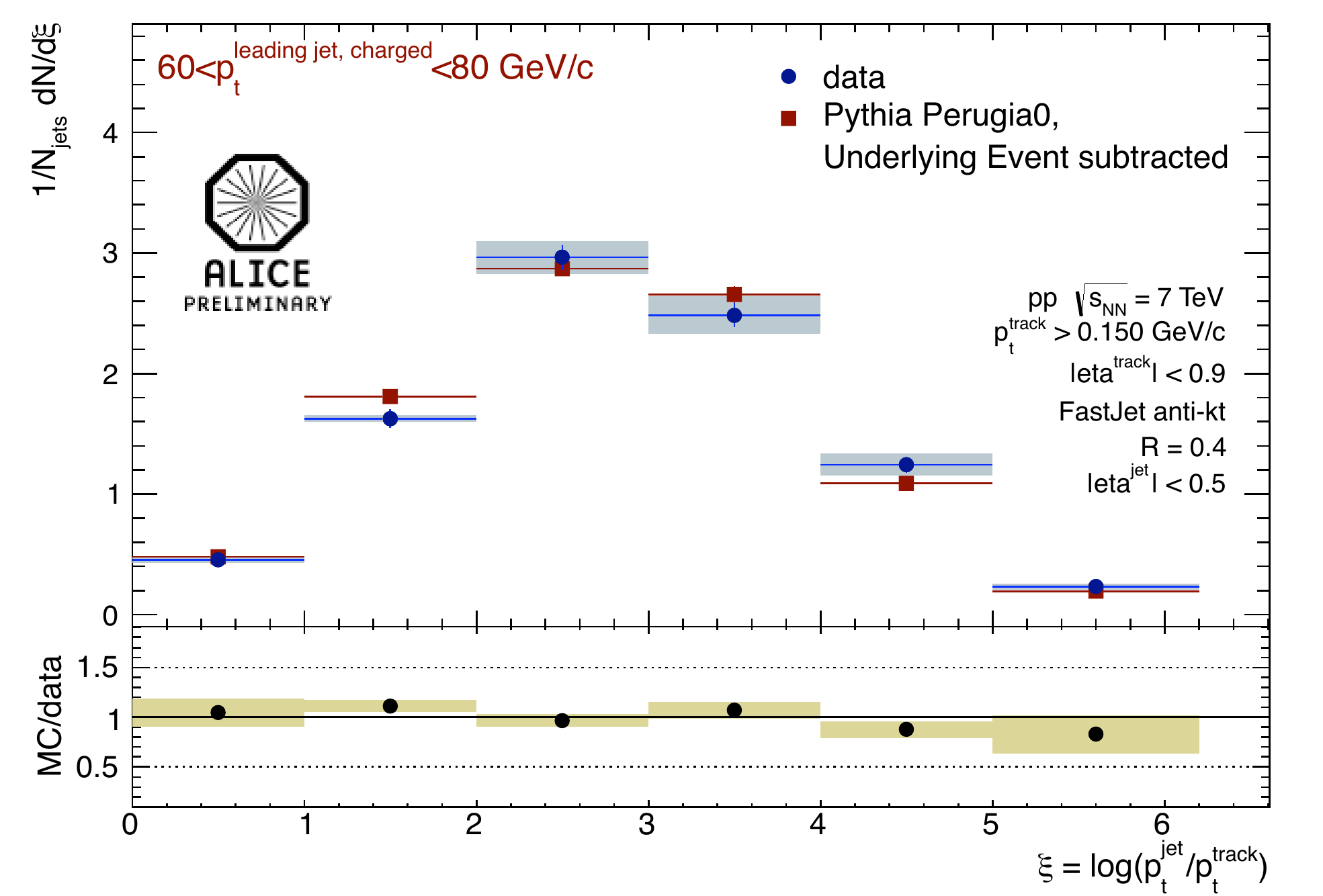}
\caption{Upper panels: \dNdxi\ distributions of charged particles within charged jets in two different charged jet \pT\ bins: 20-30 \gev\ (left) and 60-80 \gev\ (right). Data points (blue circles) are overlaid with PYTHIA6 (Perugia-0) predictions (red squares). Lower panels: ratio of PYTHIA6 predictions to data.  
\label{fig:dNdxiCH}
}
\end{figure}

Figure \ref{fig:dNdpTCH} and \ref{fig:dNdxiCH} present the \dNdpT\ and \dNdxi\ distributions of charged particles within charged jets in two different charged jet \pT\ bins: 20-30 \gev\ (left) and 60-80 \gev\ (right). Good agreement is observed between data and PYTHIA6 (Perugia-0) predictions, except at the very low \pT\ region. The hump$-$backed plateau structure \cite{HUMP} is observed in Fig. \ref{fig:dNdxiCH}, and the $\xi$ value of the peak position shifts to the right as jet \pT\ increases. 

\begin{figure}[htbp]
\centering
\includegraphics[width=0.47\textwidth]{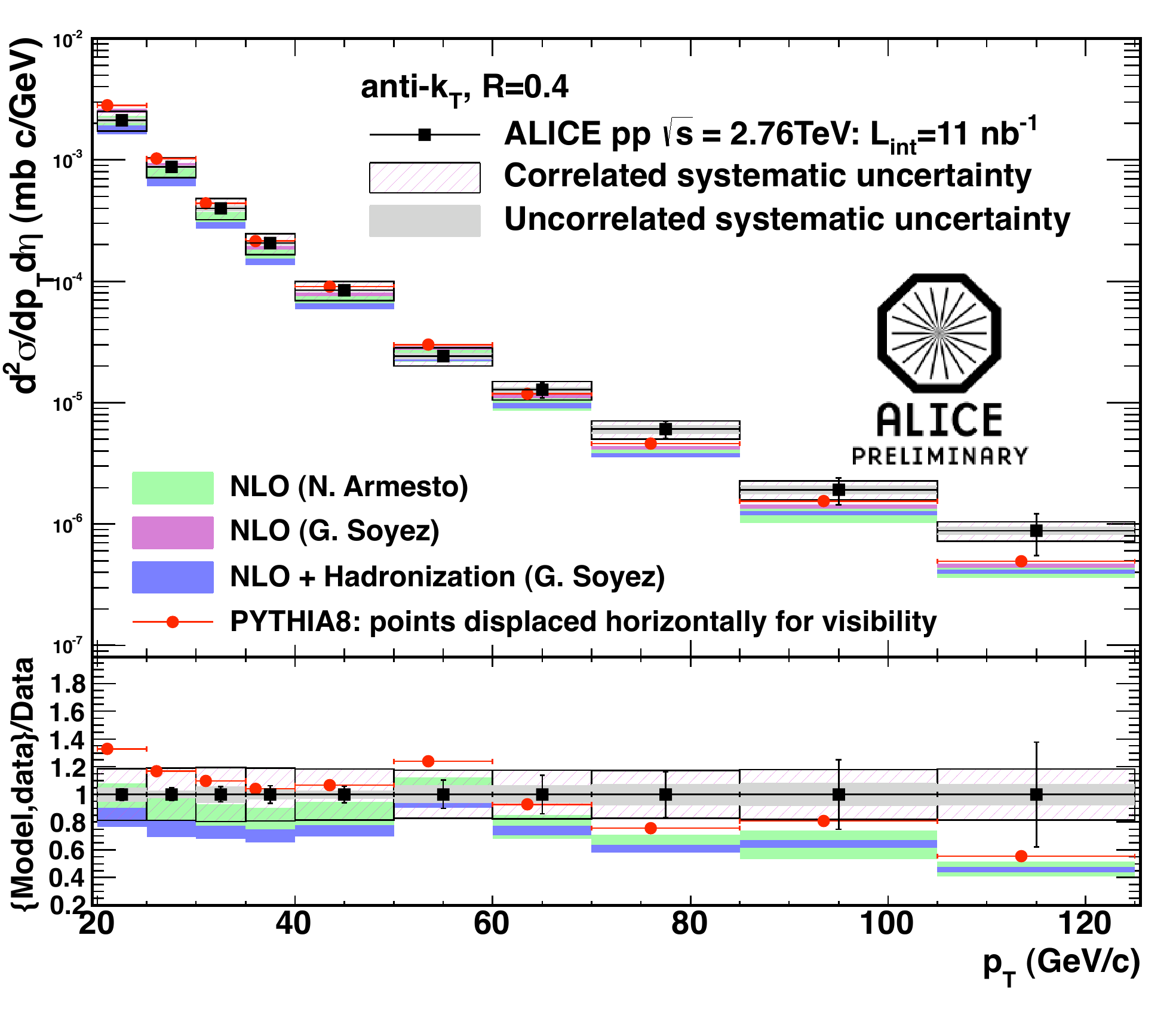}
\includegraphics[width=0.49\textwidth]{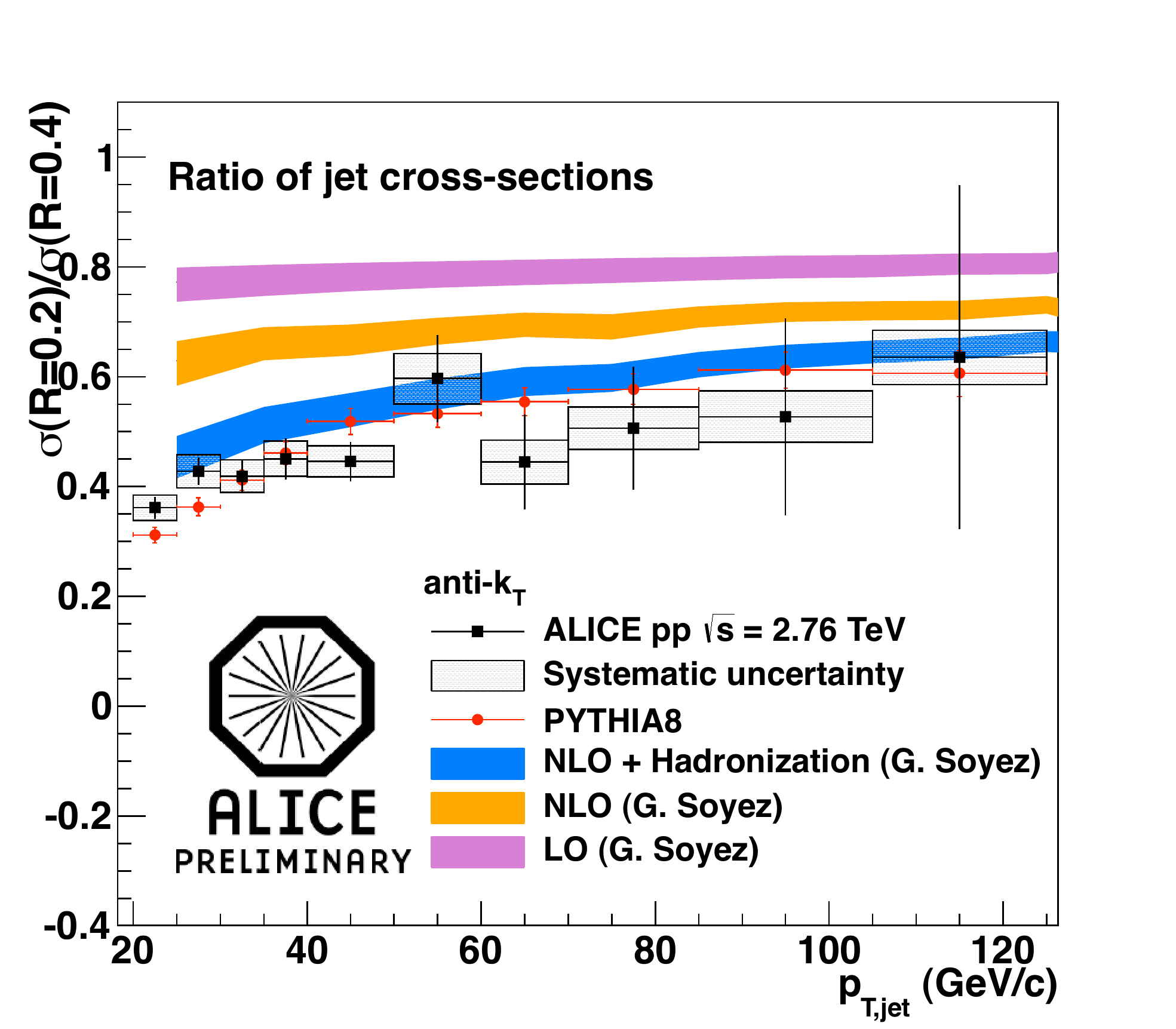}
\caption{Upper left: inclusive differential cross section obtained with \rr\ = 0.4. Data points are black squares. The colored bands show the NLO pQCD calculations, and the PYTHIA8 predictions are red circles. Lower left: ratio of NLO pQCD calculations and PYTHIA8 predictions to data. Right: ratio of inclusive differential jet cross sections for \rr\ = 0.2 and \rr\ = 0.4 with pQCD calculations and PYTHIA8 predictions.
\label{fig:jetXsec}
}
\end{figure}

Figure \ref{fig:jetXsec}, left panel, shows the inclusive differential jet cross sections obtained with \rr\ = 0.4, together with the results of the pQCD calculations at NLO \cite{SoyezRatio,Frixione:1997np} and PYTHIA8 \cite{PYTHIA8} prediction. The total uncertainties of the NLO calculations are estimated by varying the renormalisation and factorisation scales from 0.5 \pT\ to 2.0 \pT. As seen in the figure, the NLO calculations agree with data within uncertainties. The right panel of Fig. \ref{fig:jetXsec} illustrates the ratio of the inclusive jet cross sections for \rr\ = 0.2 and \rr\ = 0.4 from data, together with parton-level pQCD calculations at LO, NLO, and NLO with hadronization correction as well as PYTHIA8  predictions. The NLO calculation with hadronization correction and PYTHIA8 predictions agree with the measurement within uncertainties.

\section{Summary}
We have presented the measurements of the inclusive differential jet cross section in \pp\ collisions at \sqrts\ = 2.76 TeV, and jet fragmentation in \pp\ collisions at \sqrts\ = 7 TeV. Comparisons to theoretical calculations and various models show good agreement. Next we will carry out similar analyses in \PbPb\ collisions, and compare the results with \pp.





\bibliographystyle{elsarticle-num}
\bibliography{references}

\begin{thebibliography}{10}
\expandafter\ifx\csname url\endcsname\relax
  \def\url#1{\texttt{#1}}\fi
\expandafter\ifx\csname urlprefix\endcsname\relax\def\urlprefix{URL }\fi
\expandafter\ifx\csname href\endcsname\relax
  \def\href#1#2{#2} \def\path#1{#1}\fi

\bibitem{Majumder:2010qh}
A.~Majumder, M.~Van~Leeuwen, Prog.Part.Nucl.Phys. A66 (2011) 41--92.
\newblock \href {http://arxiv.org/abs/hep-ph/1002.2206}
  {\path{arXiv:hep-ph/1002.2206}}.

\bibitem{SoyezRatio}
G.~Soyez, PLB 698 (2011) 59.
\newblock \href {http://arxiv.org/abs/hep-ph/1101.2665v1}
  {\path{arXiv:hep-ph/1101.2665v1}}.

\bibitem{Frixione:1997np}
{S. Frixione}, Nucl.Phys. B 507 (1997) 295--314.
\newblock \href {http://arxiv.org/abs/hep-ph/9706545v1}
  {\path{arXiv:hep-ph/9706545v1}}.

\bibitem{PYTHIA}
{T. Sjostrand et al.}, JHEP 05 (2006) 026.
\newblock \href {http://arxiv.org/abs/hep-ph/0603175v2}
  {\path{arXiv:hep-ph/0603175v2}}.

\bibitem{ALICE}
{The ALICE collaboration, K. Aamodt et al.}, JINST 0803 (2008) S08002.
\newblock \href {http://arxiv.org/abs/1001.0502v3} {\path{arXiv:1001.0502v3}}.

\bibitem{Cacciari:2011ma}
{M. Cacciari, G. P. Salam and G. Soyez}, Eur.Phys.J. C72 (2012) 1896.
\newblock \href {http://arxiv.org/abs/hep-ph/1111.6097}
  {\path{arXiv:hep-ph/1111.6097}}.

\bibitem{Cacciari:2008gp}
{M. Cacciari, G. P. Salam and G. Soyez}, JHEP 04 (2008) 063.
\newblock \href {http://arxiv.org/abs/hep-ph/0802.1189}
  {\path{arXiv:hep-ph/0802.1189}}.

\bibitem{GEANT3}
{R. Brun, F. Bruyant, M. Maire, A.C. McPherson, and P. Zanarini}, {CERN Data
  Handling Division DD/EE/84-1}.

\bibitem{HUMP}
{N. Borghini and U. Wiedemann. }\href {http://arxiv.org/abs/hep-ph/0506218v1}
  {\path{arXiv:hep-ph/0506218v1}}.

\bibitem{PYTHIA8}
{T. Sjostrand et al.}, Comput.Phys.Commun 178 (2008) 852--867.
\newblock \href {http://arxiv.org/abs/hep-ph/0710.3820v1}
  {\path{arXiv:hep-ph/0710.3820v1}}.

\end{thebibliography}







\end{document}